\documentstyle[11pt,newpasp,twoside]{article}
\begin{document}

\title{QSO Absorption Lines, Gaseous Infall and Star Formation}
\author{Milan M. \'Cirkovi\'c}
%, Srdjan Samurovi\'c \& Vesna Milo\v sevi\'c-Zdjelar}}
\affil{Astronomical Observatory, Volgina 7, 11000 Belgrade,
SERBIA}
\author{Srdjan Samurovi\'c}
\affil{Department of Astronomy, University of
Trieste, Via Tiepolo 11, I-34131 Trieste, ITALY}
%\affil{ICTP, Strada Costiera 11, 34014 Trieste, ITALY}
\author{Vesna Milo\v sevi\'c-Zdjelar}
\affil{Department of Physics, University of Manitoba, Winnipeg MB,  R3T 2N2, CANADA}

\begin{abstract}
We investigate the impact of recent observational advances in research on
the gaseous content of the universe on our knowledge of star formation histories of disk galaxies.  Several contemporary surveys indicate that large quantities of gas which has not been astrated or has been astrated only weakly are bound to galaxies at later epochs ($z\le 1$).

%Identification of these clouds seen in absorption against the background QSOs
%(mainly in the form of low- and intermediate-redshift Ly$\alpha$ forest) with infalling halo
% clouds analogous to the high-velocity clouds in the Local Group offers a unique
% opportunity for building a complete picture of the baryonic dynamics in spiral galaxies.
%Such aggregates present a huge potential reservoir of gas not only for solution of the gas
%consumption puzzle in spiral disks, but also as a fuel for the future star formation.
%Thus, both the gas consumption puzzle, and problems in chemical evolution of spiral disks
% can be simultaneously solved by virtue of a single infall picture.

 \end{abstract}

%\section{Introduction}

The aim of this contribution is to investigate the impact of recent advances in
understanding of the gaseous content of the universe on our knowledge of star
formation histories of spiral galaxies. The discovery of the low-redshift
population of Ly$\alpha$ absorption systems and first steps made in understanding
of the transition between high-redshift intergalactic and low-redshift galactic
population of QSO absorbing clouds, significantly reshaped our picture of the
gaseous content of the universe. It turns out that large quantities of baryonic
matter in form of gas which has not been astrated or has been astrated only weakly
are bound to galaxies at later epochs ($z\le 1$). Such aggregates present a potential
reservoir of gas not only for solution of the gas consumption puzzle in spiral disks,
but also as a fuel for the future star formation. This baryonic transition from
diffuse toward collapsed state, although still hard to establish quantitatively, is
a result of several simple physical processes (cooling, quasi-stationary accretion, etc.),
which have only recently started to be included in viable models.

At least a fraction of the narrow absorption features in the QSO
spectra  are believed to be produced by gas in the halos of
intervening luminous galaxies of sizes $\sim$ 100 kpc (Rauch
1998), an idea going back to the classical work of Bahcall and
Spitzer (1969), and recently subjected to a plethora of both
observational and theoretical investigations (e.g. \'Cirkovi\'c, Bland-Hawthorn and Samurovi\'c 1999). Most
important of these development is the detailed redshift
coincidence analysis of a large sample of HST-observed QSO lines
of sight (Chen et al. 1998), establishing
that a large fraction ($\sim$ 60\% and maybe more) of low and
intermediate-redshift absorption lines in so-called Ly$\alpha$
forest originates in clouds residing in haloes of normal luminous
galaxies with approximately unit covering factor.

Additional lesson is that the problem of baryonic dark matter
could be solved through the hypothesis of baryonic incorporation
(Persic and Salucci 1992; Carr 1994). This problem has become more acute lately; in particular,
measurements of the strong He II Gunn-Peterson $\lambda304$
absorption seen at $\langle z \rangle \approx 3.0$ (Jakobsen
1998), imply very high diffuse baryonic density at these epochs:
$\Omega_Bh \geq 6.75 \times 10^{ -2}$ (especially if we keep in
mind recently favored low values of Hubble constant). High
baryonic density is indicated also by (newest and most precise)
observations of high-redshift deuterium (e.g.\ Burles and Tytler
1998). At low redshift, there is a massive compact object (MACHO)
component to be accounted for in the baryonic census (Fields,
Freese and Graff 1998), and careful comparative analysis is
necessary (Samurovi{\'c}, \'Cirkovi\'c and Milo\v sevi\'c-Zdjelar
1999). In particular,  uncomfortably high combined Ly$\alpha$
$+$ MACHO abundance strongly suggests the sharing the baryons
between the two major baryonic reservoirs, i.e. baryonic
incorporation.

Major virtue of the picture outlined here, in which Ly$\alpha$
clouds are located in extended haloes of luminous galaxies and
represent steadily infalling gas eventually to be turned into
stars is its capability to simultaneously solve several
different, and even historically seen as unrelated, problems in
astrophysics and cosmology. One of the main morals of the
explosive developments taking place in last two decades in both
observational and theoretical astrophysics is that there are
hardly any  isolated  problems, which could be well
treated without connection to the wider background. The gas
consumption problem in spiral disks is
%, as we have seen, 
another
instance in which we may regard present state of the universe as
somewhat ``special''. In order to avoid special initial
conditions for the (still poorly understood!) galaxy formation
process, we may choose to regard the present rate of star
formation and consumption of interstellar matter in the Milky Way
as a natural consequence of large-scale processes to which our
presence as observers is incidental, or to view it as a necessary
requisite of the growth of complexity which is to include
intelligent observers. The latter view has the
advantage of not explicitly requiring any new physical
mechanisms, and moreover, we can give a prediction on the course
of star formation history of our Galaxy.

\end{document}